\documentclass[10pt,conference]{IEEEtran}
\usepackage{silence}
\PassOptionsToPackage{dvipsnames}{xcolor}
\PassOptionsToPackage{hyphens}{url}
\PassOptionsToPackage{breaklinks,colorlinks}{hyperref}

\usepackage{amsmath,amsopn}
\makeatletter
\providecommand*\caption@documentclass{standard}
\makeatother
\usepackage{amssymb}
\usepackage{mathtools}
\usepackage{fancyhdr,lastpage}
\usepackage{times}
\usepackage[labelfont=bf,compatibility=false]{caption}
\usepackage{array,underscore,relsize}
\usepackage{microtype,xspace,graphicx,multirow}
\usepackage{booktabs}
\usepackage[T1]{fontenc}
\usepackage{enumitem}
\usepackage{pifont}
\usepackage{colortbl}
\usepackage{xcolor}
\usepackage{fancyvrb}
\usepackage{url}
\usepackage{hyperref}
\hypersetup{
  citecolor=blue,
  linkcolor=blue,
  breaklinks=true,
  urlcolor=blue
}
\usepackage[nocompress,space]{cite}
\usepackage{tabularx}
\usepackage{makecell}
\usepackage{tikz}
\usetikzlibrary{fit,calc,arrows.meta,positioning}
\usepackage{xstring}

\microtypecontext{spacing=nonfrench}

\newcommand{\sys}{\mbox{\textsc{Slyp}}\xspace}

\newcommand{\uaf}{\mbox{\textsc{UAF}}\xspace}
\newcommand{\dfree}{\mbox{\textsc{DF}}\xspace}
\newcommand{\oob}{\mbox{\textsc{OOB}}\xspace}
\newcommand{\ovf}{\mbox{\textsc{OVF}}\xspace}

\newcommand{\comrace}{\mbox{\textsc{COMRace}}\xspace}
\newcommand{\comraceplus}{\mbox{\textsc{COMRace++}}\xspace}
\newcommand{\comfusion}{\mbox{\textsc{COMFusion}}\xspace}

\newcommand{\react}{\mbox{\textsc{ReAct}}\xspace}

\newcommand{\low}{\mbox{\textsc{Low}}\xspace}
\newcommand{\medium}{\mbox{\textsc{Medium}}\xspace}

\newcommand{\system}{\mbox{\textsc{System}}\xspace}
\newcommand{\sandbox}{\mbox{\textsc{Sandbox}}\xspace}

\newcommand{\codex}{\mbox{\textsc{Codex}}\xspace}
\newcommand{\claudecode}{\mbox{\textsc{Cc}}\xspace}
\newcommand{\slyptools}{\mbox{\textsc{Slyp}} tools\xspace}
\newcommand{\TOOLS}{\mbox{Tools}\xspace}
\newcommand{\SUB}{\mbox{Sub}\xspace}
\newcommand{\syswoslyptools}{\sys (\wout{\TOOLS})\xspace}
\newcommand{\codexwithslyptools}{\codex (\with{\TOOLS})\xspace}
\newcommand{\claudecodewithslyptools}{\claudecode (\with{\TOOLS})\xspace}

\newcommand{\cc}[1]{\mbox{\smaller[0.5]\texttt{#1}}}

\fvset{fontsize=\scriptsize,xleftmargin=8pt,numbers=left,numbersep=5pt}

\input{code/fmt}
\newcommand{\figrule}{\hrule width \hsize height .33pt}
\newcommand{\coderule}{\vspace{-0.5em}\figrule\vspace{0.2em}}

\def\Snospace~{\S{}}

\makeatletter
\providecommand*{\ALG@lineautorefname}{Line}
\makeatother

\input{glyphtounicode}
\newcolumntype{R}[1]{>{\raggedleft\let\newline\\\arraybackslash\hspace{0pt}}p{#1}}

\newcommand{\squishlist}{
  \begin{itemize}[noitemsep,nolistsep]
      \setlength{\itemsep}{-0pt}
    }
\newcommand{\squishend}{
  \end{itemize}
}

\definecolor{darkgreen}{RGB}{0,100,0}
\definecolor{darkred}{RGB}{150,0,0}
\newcommand{\with}[1]{\mbox{\textcolor{darkgreen}{+#1}}}
\newcommand{\wout}[1]{\mbox{\textcolor{darkred}{-#1}}}

\newcommand*\RC[1]{%
  \begin{tikzpicture}[baseline=(C.base)]
    \node[draw,rectangle, inner sep=0.7pt](C) {#1};
\end{tikzpicture}}

\newcommand{\fullcircle}{\tikz[baseline=-0.6ex]\draw[fill=black] (0,0) circle (0.8ex);}
\newcommand{\halfcircle}{\tikz[baseline=-0.6ex]{\fill (0,0) -- (90:0.8ex) arc (90:270:0.8ex) -- cycle;\draw (0,0) circle (0.8ex);}}
\newcommand{\opencircle}{\tikz[baseline=-0.6ex]\draw (0,0) circle (0.8ex);}

\makeatletter
\newcommand{\PP}[1]{%
  \vspace{2pt}
  \noindent{\bf
    \edef\@temp{\detokenize{#1}}%
    \IfEndWith{\@temp}{.}{#1}{#1.}%
  }
}
\makeatother

\newcommand{\boxbeg}{
  \vspace{2px}
  \noindent
  \begin{tabular}{|l|}\hline
    \begin{minipage}{3.2in}
      \vspace{2px}
      \noindent
    }

    \newcommand{\boxend}{
      \vspace{2px}
    \end{minipage}\\ \hline
  \end{tabular}
  \vspace{-10pt}
}

\newcommand{\uiuc}[1]{{#1\textsuperscript{\includegraphics[scale=0.25]{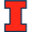}}}}

\newif\ifshowredacted
\showredactedfalse

\newlength{\redactwidth}
\newlength{\redactheight}


\begin{document}

\title{Agentic Vulnerability Reasoning on COTS Binaries}

\ifdefined\DRAFT
\pagestyle{fancyplain}
\lhead{Rev.~\therev}
\rhead{\thedate}
\cfoot{\thepage\ of \pageref{LastPage}}
\fi

\renewcommand\thefootnote{\fnsymbol{footnote}}
\IEEEoverridecommandlockouts

\author{Hwiwon Lee$^{\ast,\dagger}$\; \quad
  Jongseong Kim$^{\ast}$\; \quad
  Lingming Zhang$^{\dagger}$\;
  \\[\bigskipamount]
  \uiuc{\textit{University of Illinois Urbana-Champaign}}%
  \thanks{$^{\ast}$Equal contribution.}%
  \thanks{$^{\dagger}$Correspondence to: \cc{\{hwiwonl2, lingming\}@illinois.edu}.}
}

\date{}
\maketitle

\sloppy

\begin{abstract}

  LLM agents have been increasingly adopted for solving security tasks. However, existing evaluations usually require source code access, while commercial off-the-shelf (COTS) binaries dominate deployed software and require reasoning from stripped, optimized machine code.
  This discrepancy raises an important question: can modern LLM agents reason about vulnerabilities in critical COTS binaries? Motivated by this question, we build \sys, a \react-style pipeline for end-to-end vulnerability discovery and validation of COTS binaries.
  \sys combines extensible MCP servers for binary exploration and dynamic debugging, and validates candidate vulnerabilities by synthesizing debugger-verified proof-of-concept (PoC) crashes.
  We evaluate \sys and production coding agents, including Claude Code and Codex, on COTS Windows binaries centered on a 20-object COM benchmark.
  \sys uncovers all 64 vulnerable entry functions while default production agents miss up to 15; \sys also surfaces more true vulnerabilities than the state-of-the-art static analyzer, which discovers at most 35 with a large number of false positives.
  For validation, \sys generates debugger-verified PoCs for 67.5\% of cases, while default production agents generate none.
  Further ablations show that tool sets and model choice materially affect COTS binary reasoning. Our additional evaluation also demonstrates the generalizability of \sys on Windows kernel targets.
  To date, \sys has uncovered 39 zero-day vulnerabilities, 31 in COM/RPC services and 8 in kernel drivers, all disclosed to the Microsoft Security Response Center (MSRC), with 23 assigned CVEs and \$203{,}000 bounty awards.

\end{abstract}

\section{Introduction}
\label{s:intro}

Commercial off-the-shelf (COTS) binaries are the dominant form of software deployed in practice, and have penetrated into almost every corner of modern society.
For example, Microsoft reports that Windows runs on more than 1.4 billion monthly active devices worldwide~\cite{microsoft2025windows}.
As a result, any bug or vulnerability in such widely used systems can affect developers and end users all over the world, and contribute to the trillions of dollars in annual losses attributed to cybercrime~\cite{morgan2020cybercrime}.

While it is essential to ensure the security of COTS binaries, it is intrinsically challenging given that such binaries are typically shipped with stripped, optimized, or even obfuscated machine code without source code access~\cite{zhang2015control,kim2017cab,han2023queryx}.
For example, Windows is a high-value target for local privilege escalation, where privileged user-mode services expose Component Object Model (COM) interfaces to low-privileged clients~\cite{gu2022comrace,zhang2023comfusion} and kernel drivers expose system-call, IOCTL, and subsystem entry points that cross from sandboxed or medium-integrity code into the kernel.
Both surfaces force analysts to recover object layouts, call relationships, dispatch targets, and vulnerability-specific preconditions from decompiled code.
The challenge is not only to find a suspicious code pattern, but also to reason from binary evidence to a feasible root cause and then validate that hypothesis with a working proof-of-concept (PoC).

Traditional binary vulnerability discovery techniques have been widely studied on critical COTS targets such as Windows binaries, but they remain effective primarily within narrow target and bug-class boundaries.
Static analyzers such as \comrace~\cite{gu2022comrace} encode target-specific abstractions for Windows COM race conditions, while other systems rely on handcrafted queries, fuzzing harnesses, or source-level representations that are unavailable for production Windows binaries~\cite{jeong2019razzer,han2023queryx,zhang2023comfusion}.
These systems can produce actionable warnings, but they do not provide an end-to-end path from binary exploration to debugger-verified validation.
Manual reverse engineering remains the default way to connect decompiler output, platform metadata, execution feedback, and exploitability evidence across many entry points.

In recent years, large language models (LLMs) and AI agents have shown impressive performance across diverse domains, including security analysis~\cite{liu2024large}.
Thanks to the recent advances of frontier models, such LLM agents can autonomously invoke external tools and plan next actions to complete high-complexity, long-horizon coding tasks~\cite{yao2022react,wang2024executable}.
These agents have been evaluated on SEC-bench~\cite{lee2025secbench}, SEC-bench Pro~\cite{lee2026secbenchpro}, and CyberGym~\cite{wang2025cybergym}, and have also been reported to find a large number of zero-day vulnerabilities~\cite{projectzero2024bigsleep,projectzero2024naptime,carlini2026mythos}.
Meanwhile, existing security evaluations of modern LLM agents mainly focus on settings where they are granted full access to source code artifacts.
This discrepancy raises an important question that we aim to answer in this paper: \emph{can LLM agents effectively reason about security vulnerabilities in critical COTS binaries?}
Answering this question not only helps push the boundaries for autonomous AI security engineering, but also deepens our understanding of the intrinsic capabilities and limitations of frontier models.

To this end, we perform the first extensive study of agentic vulnerability reasoning on COTS binaries.
Besides evaluating state-of-the-art production coding agents, including \codex~\cite{openai2025codex} and Claude Code (\claudecode)~\cite{claudecode2026}, we build \sys, a standard \react-style agent pipeline targeting end-to-end vulnerability discovery and validation on COTS binaries, where validation proves a candidate vulnerability by synthesizing a debugger-verified PoC that crashes the target.
\sys supports extensible MCP tool servers covering a suite of tools for binary exploration and dynamic debugging to facilitate our security analysis.
In this study, we mainly focus on COTS Windows binaries, one of the most complicated and important COTS targets to date, and our main evaluations are performed on the widely deployed Windows COM services.

On a benchmark of 20 Windows COM objects, \sys and production agents with frontier LLMs all detect a non-trivial fraction of the studied vulnerabilities, demonstrating the power of recent LLMs on long-horizon, challenging discovery tasks over COTS binaries.
Moreover, \sys detects all 64 vulnerable entry functions while default \codex and \claudecode find 61 and 49, respectively, and it surfaces more true vulnerable functions than the state-of-the-art static analyzer, which uncovers at most 35.
For vulnerability validation, \sys synthesizes debugger-verified PoCs for 67.5\% of cases while production agents fail to produce any valid PoC.
Our study further reveals the impact of different tool sets and models on COTS binary analysis.
Lastly, we demonstrate the generalizability of \sys on Windows kernel-level COTS binaries.
In summary, the paper makes the following main contributions:
\begin{itemize}
  \item \textbf{Extensive Study.} We perform the first extensive study of frontier models and agents on vulnerability discovery for COTS binaries, and hope our study can inspire more research in this critical and promising direction.
  \item \textbf{System Implementation.} We build \sys, a standard \react-style agent pipeline for end-to-end vulnerability discovery and validation of COTS binaries. Moreover, \sys embodies a suite of extensible MCP tool servers for binary exploration and dynamic debugging, generalizable to other production agents. In total, \sys comprises 19.9K lines of Python code for its main agent framework and 8.7K additional lines for the MCP servers.
  \item \textbf{Practical Guidelines.} Our study uncovers a number of actionable guidelines that can benefit researchers and practitioners working on AI security. For example, our study shows that existing production agents can approach specialized agents for vulnerability discovery on COTS binaries, but struggle with validation through debugger-verified PoCs. Moreover, augmented with our domain-specific MCP tool servers, even production agents can perform competitive validation for COTS binaries.
  \item \textbf{Real-world Impact.} To date, this work has uncovered 39 zero-day vulnerabilities, 31 in Windows COM/RPC services and 8 in kernel drivers, all responsibly disclosed to the Microsoft Security Response Center (MSRC), with 23 assigned CVEs and \$203{,}000 in bounty awards.
\end{itemize}

\section{Background}
\label{s:bg}

In this paper, we mainly focus on COTS Windows binaries.
Existing COTS Windows vulnerability research spans multiple privilege boundaries.
In user mode, privileged service processes expose interfaces to low-privileged clients through mechanisms such as COM~\cite{microsoft2024com}.
In kernel mode, drivers expose entry points that can cross from sandboxed or medium-integrity code into \system.
Across both surfaces, source code is unavailable, compiler optimization obscures types and object layouts, and analysts must recover dispatch targets, state updates, and validation conditions from machine code.
Given that our main evaluations center on Windows COM services, the rest of this section mainly details the COM attack surface, and kernel drivers serve as the cross-surface extension evaluated in \autoref{ss:kernel-extensibility}.

\PP{COM as a Controlled Benchmark}
Windows COM is a binary interface standard for inter-process software composition~\cite{microsoft2024com}.
Hundreds of COM classes are hosted in Windows service processes that run with elevated privileges.
A low-privileged client activates a COM object by calling an activation function such as \cc{CoCreateInstance} with the target class identifier (CLSID).
The COM runtime routes the request across a process boundary via RPC over Advanced Local Procedure Call (ALPC) into the privileged service.
Two permission layers gate access.
\emph{Launch permission} controls whether the client can start the server process, and \emph{access permission} controls whether the client can invoke methods on a running object~\cite{gu2022comrace}.
Many system services grant both permissions to all authenticated users.
A memory corruption vulnerability in any reachable interface method gives the attacker code execution at the privilege level of the hosting service, making cross-process COM a dense, high-value attack surface for local privilege escalation.

\PP{Interfaces, Vtables, and the Binary Reality}
Each COM class implements one or more typed interfaces.
Interface methods are dispatched through virtual function tables (vtables) stored in the binary, where each vtable entry holds a pointer to a concrete function~\cite{gu2022comrace,zhang2023comfusion}.
Cross-process calls use a proxy/stub pair that marshals arguments via RPC between client and server~\cite{microsoft2024com}.
The Windows registry maps each CLSID to its binary path, threading model, AppID, service name, and hosting account.
In production binaries, interface methods appear as indirect call sequences through vtable pointers at fixed offsets, obscuring the interface hierarchy and concrete targets~\cite{shoshitaishvili2016sok,pawlowski2017marx}.
Recovering virtual call targets is a prerequisite for tracing data flows.

\PP{Threading Model and Root Cause of Races}
COM uses \emph{apartments} to govern concurrency~\cite{gu2022comrace}.
A single-threaded apartment (STA) serializes all incoming calls through a Windows message queue, making objects thread-safe by construction.
A multi-threaded apartment (MTA) allows concurrent method calls from multiple RPC threads, requiring objects to synchronize shared state themselves.
Most cross-process system services register as MTA for performance, so concurrent client requests execute on separate RPC threads with no framework-provided locking.
If an interface method reads or writes a member field (\cc{this+offset}) without holding a lock, the field is a data race candidate.
Conflicting accesses on the same field (read/free, write/free, free/free, read/write) produce use-after-free (\uaf), double-free (\dfree), and type confusion vulnerabilities~\cite{gu2022comrace,upadhyay2023navigating}.

\begin{figure}[!t]
  \input{code/cve-2024-49095.cpp.tex}
  \vspace{0.5\baselineskip}
  \coderule
  \caption{CVE-2024-49095: Race condition in \cc{SetPrintTicket} of PrintWorkflowUserSvc.}
  \label{fig:cve-2024-49095}
\end{figure}

\begin{figure}[!t]
  \centering
  \includegraphics[width=\columnwidth]{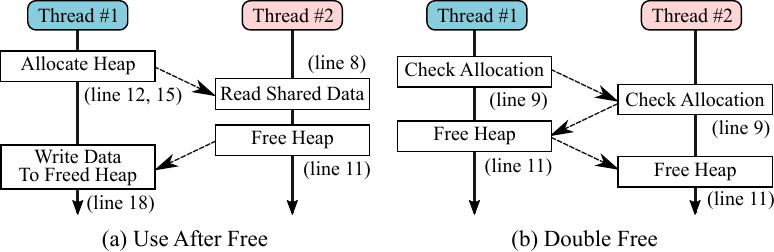}
  \caption{Two thread interleavings of concurrent \cc{SetPrintTicket} calls on the same object, producing a use-after-free (left) and a double-free (right).}
  \label{fig:cve-2024-49095-race}
\end{figure}

\PP{Running Example}
\autoref{fig:cve-2024-49095} shows the \cc{SetPrintTicket} method of the \cc{CWorkflowSession} COM class, hosted in PrintWorkflowUserSvc, a per-user service that manages the Windows Print Workflow.
The method reads the heap pointer stored at \cc{*((\_QWORD *)this + 10)} (line~8), frees the old buffer if present (line~11), allocates a new buffer (line~12), stores it back into the same field (line~15), and copies print-ticket data into it (line~18).
Since \cc{CWorkflowSession} uses the MTA threading model, concurrent RPC requests can execute this read-free-allocate-write sequence on the same object without framework-provided serialization.

\autoref{fig:cve-2024-49095-race} shows the two resulting interleavings.
In the use-after-free case, Thread~1 stores a fresh buffer in the shared field, Thread~2 reads and frees it before Thread~1 reaches \cc{memcpy}, and Thread~1 then writes into freed memory.
In the double-free case, both threads observe the same non-null pointer before either free occurs, and both invoke \cc{operator delete[]} on it.
A low-privileged client can trigger these concurrent RPC invocations, making the unsynchronized shared-field pattern security-relevant and representative of the controlled COM benchmark.
Detecting it requires COM threading semantics, cross-function shared-state tracing, and evidence that no synchronization protects the access path, which motivates \sys's combination of target-specific prompts with reusable binary exploration and debugging tools.

\section{Design}
\label{s:design}

\subsection{Overview}

\begin{figure*}[!t]
  \centering
  \includegraphics[width=\textwidth]{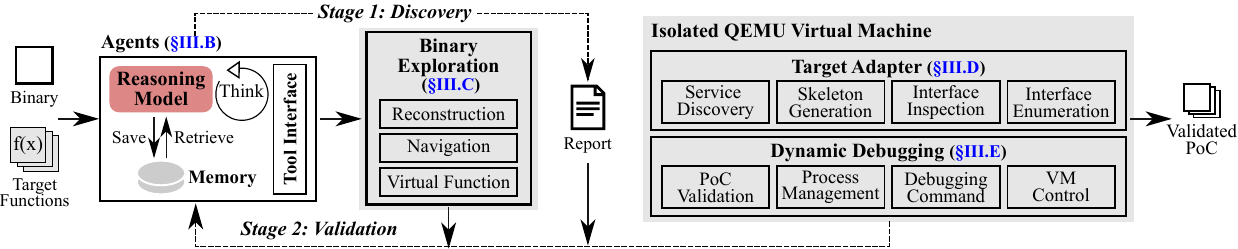}
  \caption{Overview of \sys. A single \react agent runs Stage~1 discovery and Stage~2 validation, connected by a structured vulnerability report. Target adapters provide surface-specific harness metadata when needed.}
  \label{fig:overview}
\end{figure*}

\autoref{fig:overview} illustrates the architecture of \sys, an end-to-end agentic pipeline for vulnerability reasoning on COTS binaries.
Given a binary, a set of exposed entry points, and a vulnerability-class specification, \sys determines whether a vulnerability is reachable, localizes its root cause in decompiled code, and generates a PoC that triggers debugger-observable evidence such as access violation errors. The vulnerability-class specification supplies target semantics such as synchronization APIs, object lifetime rules, and valid call sequences, and keeping these semantics outside the core binary exploration and debugging tools lets new targets be approached through prompt and adapter changes rather than new analyzers.

The pipeline operates in two stages connected by a structured vulnerability report.
In Stage~1 (Vulnerability Discovery), an LLM agent uses binary exploration tools to navigate decompiled code, resolve dispatch, trace data and control flow, and identify vulnerability hypotheses across target entry functions.
In Stage~2 (Validation), the agent uses dynamic debugging tools to compile, execute, and debug C/C++ PoC programs against a live Windows target, iteratively refining until a debugger-verified crash is captured.
Target-specific adapters are optional. Ours is instantiated for COM to supply activation metadata and skeleton code for the COM benchmark, while kernel targets bypass it and reuse the same binary exploration and debugging cores.

\sys exposes three categories of MCP tool interfaces.
\textbf{Binary exploration} (\autoref{ss:binary}) provides decompilation with integrated virtual dispatch resolution, disassembly, cross-references, and call graph navigation for static binary analysis.
\textbf{Target adapters} expose platform-specific activation or harness metadata when a target requires it. Our adapter, instantiated for COM (\autoref{ss:com}), provides COM registry enumeration, interface declarations, security descriptors, and skeleton code generation.
\textbf{Dynamic debugging} (\autoref{ss:poc}) supports PoC compilation, execution, crash capture, and heap diagnostics against a live Windows environment.
All tools are accessible through the Model Context Protocol (MCP)~\cite{anthropic2024mcp}, enabling any MCP-compatible agent to invoke the same interfaces.
The pipeline is model-agnostic and target-adaptable. Changing the vulnerability class or attack surface changes the task prompt and optional adapter usage, not the Stage~1 or Stage~2 cores.

\subsection{Agent Architecture}
\label{ss:agent}

\sys implements a standard \react agent~\cite{yao2022react}, the same think-act-observe paradigm used by production coding agents such as \codex~\cite{codex2025} and \claudecode~\cite{claudecode2026}.
At each step, the model generates an action, either a tool call or a final response, and the tool interface dispatches the call to the appropriate MCP server or built-in tool and returns an observation that the model reasons over before selecting the next action.
The loop terminates when the agent produces a structured vulnerability report or exhausts the step budget.
A middleware pipeline runs before each turn and before the agent stops, enabling the context management and verification behaviors described below.
The differences between agents lie in specific engineering choices within this shared framework, not in a fundamentally different architecture.

\PP{Context Management}
Long-horizon binary analysis generates large trajectories, since a single decompilation produces thousands of tokens and examining dozens of entry functions with their callees generates tons of tokens.
\sys addresses this growth through auto-compaction with structured checkpointing, triggered by a middleware monitor as token usage approaches the context window limit.
During compaction, the system preserves recent tool-call/response pairs, saves substantive assistant messages to an external checkpoint file as a lossless side channel, and asks a compaction model to generate a structured summary of the task, methodology, findings with evidence, cross-function relationships, coverage, external state, and current work state.
A post-compaction directive then instructs the agent to read the saved checkpoint, re-verify tentative conclusions, and continue analysis from the recovered state.
If a single tool output exceeds the context window before compaction triggers, the system applies progressive truncation at three levels until the context fits.

\PP{Memory}
\sys provides a file-based memory system organized as a compact index plus topic files. The index loads into the agent's context at each turn and provides section headers, inline findings, and links to detailed topic files. Each topic file stores one durable analytical finding with YAML frontmatter for its name, description, and type, and the agent incrementally saves function summaries, vulnerability patterns, and cross-function relationships there so they survive compaction. Retrieval uses deterministic BM25 search~\cite{robertson2009bm25} augmented with frontmatter text, which lets the agent recover specific evidence after compaction without extra model calls or full-history rereading. This design keeps the active context bounded while detailed evidence accumulates on disk.

\PP{Task Verification}
We use verifier-style middleware to keep the agent aligned with the task throughout long analysis sessions. Instead of trusting the agent's self-assessment, the middleware inspects observable execution state and injects corrective feedback when the trajectory shows that the task has not been carried out or no usable finding has been produced. A stop-time verifier prevents premature termination by rejecting plan-only exits, requiring actual tool use in task-oriented sessions, and catching empty final turns after tool execution until the agent emits a usable artifact. During long trajectories, a token-budget-driven memory nudge asks the agent to consolidate durable findings into external memory and re-check that the current line of work still matches the original objective, preserving analytical state during tool-heavy phases without disrupting shorter reasoning phases.

\subsection{Binary Exploration}
\label{ss:binary}

Analyzing compiled binaries requires specialized tools that differ fundamentally from source-code analysis.
Even with debug symbols, binaries lack the full semantic context available in source, so object layouts, type hierarchies, calling conventions, and inter-procedural data flows must be recovered through decompilation and cross-reference analysis.
Any COTS surface requires this recovery step, with the specific dispatch and data flow structures varying by target. COM services rely heavily on virtual dispatch through vtables, whereas kernel drivers require navigation across dispatch routines, helper functions, and subsystem callbacks.
We implement a suite of binary exploration tools through MCP~\cite{anthropic2024mcp}, providing a unified interface for vulnerability analysis over compiled code.

\autoref{tab:mcp-tools} lists the agent-facing tools exposed during evaluation.
Code reconstruction tools (\cc{decompile\_function\_addr/name}, \cc{disassemble\_function}) transform binary code into high-level pseudocode with integrated virtual function call resolution or instruction-level assembly.
\cc{get\_function\_by\_name} and \cc{get\_function\_by\_address} retrieve function metadata including bounds and argument arity.
Navigation tools enable tracing how data and control flow propagate through call chains.
\cc{get\_xrefs\_to} identifies all code locations referencing a given address, enabling agents to locate all uses of shared state or calls to security-relevant functions.
\cc{get\_callees} and \cc{get\_callers} enumerate direct call relationships, and \cc{search\_functions} finds matching functions by name pattern.
Function addresses serve as shared identifiers across code reconstruction and successive call-graph navigation steps.

Virtual function calls in C++ binaries, including COM services, appear as indirect calls through vtable pointers, obscuring the actual target function~\cite{pawlowski2017marx,schwartz2018ooanalyzer}.
We integrate virtual function call resolution directly into the decompilation process, automatically resolving virtual calls and annotating them in the pseudocode output.
For each indirect call, the system identifies whether it represents a virtual call by analyzing the instruction pattern.
An object tracer then follows the object backward through its definition chain, collecting candidate vtables by analyzing cross-references and constructor invocations.
The system parses the method offset, looks up the concrete function in each candidate vtable, and annotates the pseudocode with resolved function names and addresses.
This eliminates the need for agents to arduously reason about virtual dispatch and ensures that resolved targets appear as concrete callees in navigation results.
Vtables loaded dynamically from external binaries and polymorphic initialization from multiple classes remain unresolved by this static technique.

\begin{table}[t]
  \centering
  \caption{MCP tools provided to the agent across three servers: binary exploration (9 tools), target adapter (14 tools), and dynamic debugging (10 tools).}
  \label{tab:mcp-tools}
  \footnotesize
  
\begin{tabular}{@{}lp{5.2cm}@{}}
  \toprule
  \textbf{Tool}                   & \textbf{Description} \\
  \midrule
  \rowcolor[RGB]{234, 234, 234}\multicolumn{2}{@{}c}{\textit{Binary Exploration}} \\
  \cc{decompile\_function\_addr}  & Pseudocode with vtable resolution (by address) \\
  \cc{decompile\_function\_name}  & Pseudocode with vtable resolution (by name) \\
  \cc{disassemble\_function}      & Instruction-level assembly \\
  \cc{get\_function\_by\_name}    & Function metadata by symbol name \\
  \cc{get\_function\_by\_address} & Function metadata by address \\
  \cc{search\_functions}          & Pattern-based function search \\
  \cc{get\_xrefs\_to}             & Cross-references to an address \\
  \cc{get\_callees}               & Direct callees of a function \\
  \cc{get\_callers}               & All callers of a function \\
  \addlinespace
  \rowcolor[RGB]{234, 234, 234}\multicolumn{2}{@{}c}{\textit{Target Adapter}} \\
  \cc{find\_service}              & Locate COM service by name \\
  \cc{list\_runtime\_servers}     & Enumerate WinRT runtime servers \\
  \cc{list\_local\_services}      & Enumerate classic COM local services \\
  \cc{get\_server\_classes}       & Classes hosted by a runtime server \\
  \cc{get\_class\_interfaces}     & Interfaces of a WinRT class \\
  \cc{get\_service\_clsids}       & CLSIDs registered to a service \\
  \cc{get\_clsid\_info}           & CLSID metadata and threading model \\
  \cc{get\_clsid\_interfaces}     & Interfaces of a classic COM class \\
  \cc{get\_security\_descriptor}  & Launch/access permissions (SDDL) \\
  \cc{find\_interface}            & Look up interface by name or IID \\
  \cc{get\_interface\_info}       & Interface metadata and vtable offset \\
  \cc{get\_proxy\_definition}     & Method signatures from proxy/stub DLL \\
  \cc{generate\_skeleton\_code}   & Compilable C++ activation template \\
  \cc{test\_oleview\_connection}  & Verify server connectivity \\
  \addlinespace
  \rowcolor[RGB]{234, 234, 234}\multicolumn{2}{@{}c}{\textit{Dynamic Debugging}} \\
  \cc{submit\_poc}                & Compile, execute, and capture crash \\
  \cc{run\_command}               & Execute a debugger command \\
  \cc{run\_sequence}              & Execute a command sequence \\
  \cc{get\_process\_list}         & List running processes \\
  \cc{list\_service\_pids}        & Enumerate service process IDs \\
  \cc{start\_service}             & Start a Windows service \\
  \cc{attach\_process}            & Attach debugger to a process \\
  \cc{detach\_process}            & Detach debugger from a process \\
  \cc{force\_break}               & Break into the debugger \\
  \cc{resume\_vm}                 & Resume VM execution \\
  \bottomrule
\end{tabular}

\end{table}

\subsection{Target Adapter}
\label{ss:com}

COTS binaries can require activation or harness metadata that is difficult to recover from the target binary alone.
For example, for COM, writing a PoC requires knowing how to activate the target object, which CLSID and interfaces to use, what method signatures look like, and what privilege level the service runs at.
This information resides in the Windows registry and in proxy/stub DLLs, not in the target binary itself.
We expose capabilities such as COM registry inspection as dynamic MCP interfaces, giving the LLM agent on-demand access to live metadata.
Such a target adapter is invoked only when a surface requires activation or harness metadata that is hard to infer from the binary alone.
In this paper, the target adapter is instantiated only for COM, where it is essential for automated COM harness generation, while kernel drivers bypass it entirely.
The same target adapter can also be extended to support other COTS binaries that require additional metadata.

The tool interface provides 14 tools organized by workflow stage (\autoref{tab:mcp-tools}).
\emph{Service discovery} locates target COM services and their hosting model.
\emph{Class and interface enumeration} retrieves classes, interfaces, and metadata (threading model, privilege level, AppContainer accessibility) with separate paths for WinRT and classic COM CLSIDs.
\emph{Security analysis} retrieves launch and access permissions in SDDL format to determine whether an unprivileged client can activate the target.
\emph{Interface inspection} retrieves full method signatures in Interface Definition Language (IDL) or C++ format from proxy/stub DLLs.
\emph{Code generation} produces compilable C++ skeletons with the correct activation pattern (\cc{CoCreateInstance} for classic COM, \cc{RoActivateInstance} or \cc{RoGetActivationFactory} for WinRT) working as a placeholder for crafting a PoC.

\subsection{Dynamic Debugging}
\label{ss:poc}
Stage~1 produces candidate vulnerabilities that can include hallucinations or static false positives, so they require crash evidence to reject unfounded reports and confirm true positives.
Stage~2 addresses this with automated PoC generation paired with debugger-based crash verification, which also yields concrete artifacts such as source code, crash dumps, call stacks, and register states that support triage and disclosure.
The loop is shared across targets, while the harness code differs by target.
COM PoCs require activation and concurrent interface invocations, whereas kernel PoCs use driver-style interaction patterns such as device handles and IOCTLs.

The PoC pipeline receives Stage~1's vulnerability report as input.
The agent extracts the affected component, entry functions, vulnerability type, and suggested trigger strategy from the structured report.
For COM cases, it uses the target adapter (\autoref{ss:com}) to retrieve interface definitions and generate a skeleton C++ template, then writes the concurrent invocation logic to trigger the identified race window.
For kernel cases, it skips the target adapter and writes a driver-style harness while reusing the same compilation, execution, and debugging loop.
The agent enters an iterative compile-execute-debug loop~\cite{xia2024automated}, submitting PoC source code, receiving compilation errors or execution results (crash dump, timeout, or clean exit), and refining the source based on the feedback.
We enable page heap on the target VM so that race-induced heap corruptions are caught deterministically at the faulting instruction.
A PoC is verified when the crash dump shows an access violation, heap corruption, security check failure, or debugger break and the faulting stack remains attributable to the Stage~1 target.

The tool interface provides 10 debugging tools in four categories.
\emph{PoC submission} compiles C++ source on the target VM and captures output and crash evidence in a single tool call.
\emph{Command execution} tools run arbitrary debugger commands and command sequences for targeted analysis.
\emph{Process and service management} tools enumerate running processes, start services, and attach or detach the debugger.
\emph{VM control} tools break into the debugger and resume VM execution when the target service requires recovery.

\section{Implementation}
\label{s:impl}
We implement \sys as a 19.9K-line Python agent framework plus 8.7K lines across three MCP servers. The reusable core consists of binary exploration and dynamic debugging servers, while the optional target adapter supplies user-mode metadata and activation. The binary exploration server exposes 9 tools on top of IDA Pro~\cite{idapro}: a Python FastMCP server drives an IDA plugin over HTTP JSON-RPC, auto-generates tool wrappers from plugin decorators, implements vtable resolution and callee/caller enumeration atop IDA Hex-Rays, and uses Microsoft public debug symbols.

The target adapter exposes 14 OleViewDotNet-backed tools~\cite{oleviewdotnet,gu2022comrace} through a host FastMCP server connected to a C\# .NET application inside the target VM over a TCP JSON-line protocol. Kernel targets bypass this adapter and reuse the binary exploration and debugging servers with prompt-level instructions. The debugging server exposes 10 tools through FastMCP by driving a C++ WinDbg extension DLL over a named pipe, using the WinDbg engine API for user-mode and kernel debugging. PoCs compile and run on the target VM through Windows Remote Management and MSVC, and all evaluation runs use a QEMU-hosted Windows~11 VM with an attached debugger and per-case snapshots for isolation. The agent framework uses async generator-based streaming with before-turn and before-stop middleware hooks for compaction, memory nudging, and task verification (\autoref{ss:agent}), and a shared MCP registry caches tool discovery so \codex, \claudecode, and \sys use the same servers unmodified.

\section{Evaluation}
\label{s:eval}

We evaluate \sys to answer four research questions:
\begin{itemize}[leftmargin=1.5em]
  \item \textbf{RQ1.} How does \sys compare with production coding agents on COM vulnerability discovery and validation under their default configurations?
  \item \textbf{RQ2.} How do different toolsets and models contribute to the overall performance of \sys?
  \item \textbf{RQ3.} How does \sys compare to traditional static COM race-condition analysis?
  \item \textbf{RQ4.} Beyond the controlled benchmark, what does \sys discover on additional COM services and on a different kernel attack surface, reusing the same pipeline?
\end{itemize}

We present \sys as both a hybrid system and an empirical contribution: a from-scratch \react scaffold augmented with tailored tools, evaluated against production agents and a static analysis system on a controlled benchmark, and then tested for generalization beyond it.

\subsection{Experimental Setup}
\label{ss:experimental-setup}

We separate vulnerability reasoning into \emph{discovery} and \emph{validation}: discovery locates candidate vulnerable functions, while validation proves a candidate by synthesizing a PoC whose debugger-verified crash demonstrates the finding.
We study race-induced memory-corruption \emph{vulnerabilities}, specifically \uaf and \dfree conditions, and confirm them by \emph{debugger-verified} memory access violations.
A vulnerability is the memory-safety defect and a PoC triggers it and produces an observable crash.
We measure discovery by recovered vulnerable entry functions and validation by cases with debugger-verified crashes from agent-generated PoCs.
We therefore report discovery and validation separately, and use the local privilege escalation (LPE) security boundaries in \autoref{tab:benchmark-0days}.

\subsubsection{Dataset}
\label{sss:dataset}
For RQ1--RQ3, we assemble a controlled benchmark of 20 production Windows COM objects across ten services.
We prioritize previously unexplored services reachable from \low-privilege clients to avoid rediscovering patched bugs and to exercise realistic LPE surfaces.
Within each selected service, we include every COM object that an authenticated user can activate and whose interface metadata we can recover, so the selection is exhaustive within scope.
The benchmark targets \uaf and \dfree conditions arising from COM race conditions.

We label ground truth through a uniform validation pipeline, not from any single agent's output.
All three agents (\sys, \codex, \claudecode) analyze the same binaries under identical conditions, and we manually validate every surfaced candidate, together with the original vulnerability reports, by building a crashing PoC.
An entry function is labeled vulnerable only when \ding{182} a race occurs in the function or its callees, or \ding{183} it participates in a vulnerable setter/getter interaction, and \ding{184} the location is reachable by a PoC that crashes the service, with final confirmation from MSRC.

This procedure yields 379 entry functions across the 20 objects, of which 64 are confirmed vulnerable, with per-object complexity ranging from 3 to 69 entry functions.
Twelve objects carry 24 MSRC-confirmed zero-day instances detected by our study (\autoref{tab:benchmark-0days}), and the remaining eight are 1-day objects reported independently by other researchers.
The ground truth is therefore the union of every vulnerable function validated from candidates surfaced by \sys, \codex, and \claudecode, plus the independent 1-day reports, not only \sys's findings.
We report the number of vulnerable functions recovered and the false-positive count against this manually constructed, MSRC-confirmed ground truth.
We use these 64 vulnerable functions as the discovery ground truth.
For PoC generation, we instead group the 32 vulnerabilities into 40 cases, because one vulnerability's fix can span multiple functions and a race triggers per interacting function set rather than per function.
Each case is a minimal race pattern completed by one self-racing entry function or by a set of functions whose concurrent invocations realize the interleaving.

\subsubsection{Experimental Protocol}
\label{sss:protocol}
For RQ1--RQ3, we evaluate all agents on the full 20-object COM benchmark under two controls: \ding{182}~tool access through standardized MCP interfaces for binary exploration, the target adapter, and dynamic debugging, with default or full-tool configurations defined per RQ, and \ding{183}~identical task prompts for race-induced memory-corruption discovery.
We score discovery over the 64 vulnerable entry functions and validation over the 40 vulnerability cases.

For discovery, we execute three independent runs per agent configuration and report each object's strongest run, pooled across all 20 objects, as discovery has no validation feedback and is sensitive to nondeterministic analysis coverage and reporting.
For validation, each case receives a scoped input of one or two vulnerable functions, and the dynamic debugging toolset provides compilation, execution, and crash feedback for iterative PoC repair.
RQ4 explores generalization of \sys, additional zero-day findings beyond the benchmark, and required engineering changes for extension.

\subsubsection{Baselines}
\label{sss:baselines}
We compare \sys against two state-of-the-art production-level coding agents: OpenAI Codex CLI (\codex)~\cite{openai2025codex} on GPT-5.4 and Anthropic Claude Code (\claudecode)~\cite{claudecode2026} on Opus~4.6.
Both agents invoke tools autonomously in a \react-style loop~\cite{yao2022react} and support context compaction.
In RQ1, production agents use their default binary analysis support, approximated by the minimal binary analysis configuration available to \codex and \claudecode.
In RQ2, they connect to the same MCP servers as \sys, with toolset configurations specified per experiment.
There are no limits on budget or reasoning steps.
Same-model pairing controls for model capability: \sys and \codex share GPT-5.4, and \sys and \claudecode share Opus~4.6 through the same API endpoints.
For model-generalization in RQ2, we run \sys on four additional models (GPT-5.3-codex, GPT-5.4-mini, Sonnet~4.6, Haiku~4.5) spanning capability and cost tiers.
RQ3 compares \sys with \comrace~\cite{gu2022comrace}, a traditional static binary analysis tool for detecting race conditions in COM services.
Due to limited availability of \comrace, we reproduce \comrace and its extended version \comraceplus from scratch with improved vtable recovery logic, implementing \comraceplus.

\subsection{RQ1: Discovery and Validation vs. Production Agents}
\label{ss:default-comparison}

We compare \sys against default \codex and \claudecode on both Stage~1 vulnerability discovery and Stage~2 validation, using two frontier agent-model pairs (\codex on GPT-5.4, \claudecode on Opus~4.6).
The default production-agent setting is approximated by the minimal binary analysis configuration (decompilation, disassembly, and symbol lookup via \cc{get\_function\_by\_name} and \cc{get\_function\_by\_address}), which is the most direct binary-navigation support available in the baseline setting.
Discovery is evaluated at the granularity of 20 COM objects and validation at the granularity of 40 vulnerability cases, where each case receives a scoped Stage~1 report and the agent writes, compiles, and executes C++ PoC code against a live COM service in a QEMU-hosted Windows~11 VM with \cc{cdb.exe} attached through the WinDbg MCP server.

\begin{table*}[!t]
  \centering
  \caption{Baseline performance on two frontier pairs. \textbf{Found} counts detected vulnerable entry functions out of 64, \textbf{FP} counts false positives, and \textbf{FP Rate} is $\mathrm{FP}/(\mathrm{TP}+\mathrm{FP})$, pooled across each object's strongest run. \textbf{Succ.} counts cases with a verified PoC out of 40, and \textbf{\# PoCs} counts submitted PoCs. BIN, ADP, and DBG denote binary exploration, target adapter, and dynamic debugging tools. Tool-call, token, and runtime columns are averages. In. Tok is millions, Out. Tok is thousands.}
  \resizebox{\textwidth}{!}{
    \setlength{\tabcolsep}{2.0pt}
\begin{tabular}{@{}ll rrrrccc rrrrcrrrr@{}}
  \toprule
  \multirow{2}{*}{\textbf{Model}} & \multirow{2}{*}{\textbf{Agent}}
                                  & \multicolumn{7}{c}{\textbf{Vulnerability Discovery}}
                                  & \multicolumn{9}{c}{\textbf{Vulnerability Validation}} \\
  \cmidrule(lr){3-9}\cmidrule(l){10-18}
                                  &                                          & \makecell{\textbf{Tool}\\\textbf{Calls}} & \makecell{\textbf{In.}\\\textbf{Tok}} & \makecell{\textbf{Out.}\\\textbf{Tok}} & \textbf{Runtime}
                                  & \makecell{\textbf{Found}\\\textbf{/64}}  & \textbf{FP}                              & \makecell{\textbf{FP}\\\textbf{Rate}}
                                  & \makecell{\textbf{Tool}\\\textbf{Calls}} & \textbf{\# PoCs}                         & \textbf{BIN}                          & \textbf{ADP}                           & \textbf{DBG} & \makecell{\textbf{In.}\\\textbf{Tok}} & \makecell{\textbf{Out.}\\\textbf{Tok}} & \textbf{Runtime} & \textbf{Succ.} \\
  \midrule
  \multirow{2}{*}{GPT-5.4}
                                  & \cellcolor{gray!15}\sys
                                  & \cellcolor{gray!15}111.7
                                  & \cellcolor{gray!15}2.9M
                                  & \cellcolor{gray!15}98K
                                  & \cellcolor{gray!15}34.8m
                                  & \cellcolor{gray!15}\textbf{64/64}
                                  & \cellcolor{gray!15}\textbf{2}
                                  & \cellcolor{gray!15}\textbf{3.0\%}
                                  & \cellcolor{gray!15}109.7
                                  & \cellcolor{gray!15}129
                                  & \cellcolor{gray!15}44.5
                                  & \cellcolor{gray!15}16.4
                                  & \cellcolor{gray!15}48.9
                                  & \cellcolor{gray!15}5.6M
                                  & \cellcolor{gray!15}88K
                                  & \cellcolor{gray!15}43.3m
                                  & \cellcolor{gray!15}\textbf{23/40} \\
                                  & \codex
                                  & 67.7
                                  & 1.0M
                                  & 11K
                                  & 4.4m
                                  & 61/64
                                  & 5
                                  & 7.6\%
                                  & 29.0
                                  & 41
                                  & 29.0
                                  & 0.0
                                  & 0.0
                                  & 0.9M
                                  & 14K
                                  & 4.8m
                                  & 0/40 \\
  \midrule
  \multirow{2}{*}{Opus 4.6}
                                  & \cellcolor{gray!15}\sys
                                  & \cellcolor{gray!15}58.5
                                  & \cellcolor{gray!15}3.3M
                                  & \cellcolor{gray!15}37K
                                  & \cellcolor{gray!15}13.8m
                                  & \cellcolor{gray!15}\textbf{64/64}
                                  & \cellcolor{gray!15}\textbf{4}
                                  & \cellcolor{gray!15}\textbf{5.9\%}
                                  & \cellcolor{gray!15}91.0
                                  & \cellcolor{gray!15}61
                                  & \cellcolor{gray!15}22.6
                                  & \cellcolor{gray!15}8.0
                                  & \cellcolor{gray!15}59.9
                                  & \cellcolor{gray!15}7.7M
                                  & \cellcolor{gray!15}67K
                                  & \cellcolor{gray!15}49.6m
                                  & \cellcolor{gray!15}\textbf{27/40} \\
                                  & \claudecode
                                  & 50.5
                                  & 3.7M
                                  & 130K
                                  & 12.6m
                                  & 49/64
                                  & 17
                                  & 25.8\%
                                  & 55.6
                                  & 51
                                  & 55.6
                                  & 0.0
                                  & 0.0
                                  & 6.6M
                                  & 217K
                                  & 21.8m
                                  & 0/40 \\
  \bottomrule
\end{tabular}

  }
  \label{tab:rq1-baseline}
\end{table*}

\PP{Discovery}
For vulnerability discovery, as detailed in \autoref{tab:rq1-baseline}, \sys and existing production agents can all detect a non-trivial number of vulnerable entry functions (49 to 64) with a reasonable number of false positives (2 to 17), demonstrating the promising future of LLM agents for COTS binary vulnerability discovery.
In addition, the domain-specific \sys outperforms both production agents, recovering all 64 vulnerable entry functions with both GPT-5.4 and Opus~4.6, while default \codex finds 61 and default \claudecode finds only 49.
\codex and \claudecode also produce 2.5$\times$ and 4.3$\times$ as many false positives as \sys, respectively.
The main reason is that production agents cannot reconstruct dispatch relationships from raw decompilation without structured cross-reference and vtable navigation, and thus both miss potential vulnerable functions and over-report benign ones.
In contrast, \sys's longer discovery sessions reflect coverage-driven verification.
On GPT-5.4, \sys spends the extra time to reach full coverage while the faster \codex run leaves vulnerable functions uncovered.
On Opus~4.6, \sys and \claudecode spend comparable time, yet \sys recovers every vulnerable function while \claudecode misses 15, which separates elapsed time from analysis capability.

\PP{Validation}
Surprisingly, while \sys remains competitive for validation (23/40 with GPT-5.4 and 27/40 with Opus~4.6), production agents entirely collapse and cannot produce a single valid PoC.
Without the adapter's activation metadata, production agents cannot construct the CLSID/IID initialization sequences that COM harnesses require, and without execution feedback they cannot compile, run, or diagnose a PoC at all.
The 0/40 result is structural, not model-specific, since it holds for both frontier backbones, identifying execution feedback as the minimum requirement for vulnerability validation.
We also observe that \claudecode spends 4.5$\times$ more runtime than \codex in this setting, yet both verify no cases.
This contrast shows that additional wall-clock time does not compensate for missing COM activation metadata and debugger feedback.
It also motivates our next research question on the impact of tool support for agentic security reasoning on COTS binaries.

\begin{table}[!t]
  \centering
  \caption{Zero-day vulnerabilities in the controlled COM benchmark evaluated in RQ1--RQ3. The table lists \protect\RC{01}--\protect\RC{24}, covering 12 of 20 objects across 6 of 10 services. ``Merged'' indicates a shared patch with a separate bounty. ``Confirmed'' indicates MSRC acknowledgement with a bounty award.}
  \resizebox{\columnwidth}{!}{%
    \setlength{\tabcolsep}{3pt}
{
  \begin{tabular}{c c c c c}
    \toprule
    \textbf{\#} & \textbf{Service}                                            & \textbf{Type} & \textbf{Sec.~Boundary}     & \textbf{Status} \\
    \midrule
    \RC{01}     & \multirow{3}{*}{\makecell[c]{DeviceAssociation\\BrokerSvc}} & \uaf          & \low~\ding{233}~\medium    & CVE-2025-50174 \\
    \RC{02}     &                                                             & \dfree        & \low~\ding{233}~\medium    & Merged \\
    \RC{03}     &                                                             & \dfree        & \low~\ding{233}~\medium    & Merged \\
    \midrule
    \RC{04}     & \multirow{8}{*}{\makecell[c]{PrintWorkflow\\UserSvc}}       & \uaf          & \low~\ding{233}~\medium    & CVE-2025-53133 \\
    \RC{05}     &                                                             & \uaf          & \low~\ding{233}~\medium    & CVE-2025-55688 \\
    \RC{06}     &                                                             & \uaf          & \low~\ding{233}~\medium    & CVE-2025-55684 \\
    \RC{07}     &                                                             & \uaf          & \low~\ding{233}~\medium    & CVE-2025-55691 \\
    \RC{08}     &                                                             & \uaf          & \low~\ding{233}~\medium    & CVE-2025-55690 \\
    \RC{09}     &                                                             & \uaf          & \low~\ding{233}~\medium    & CVE-2025-55689 \\
    \RC{10}     &                                                             & \uaf          & \low~\ding{233}~\medium    & CVE-2025-55686 \\
    \RC{11}     &                                                             & \uaf          & \low~\ding{233}~\medium    & CVE-2025-55685 \\
    \midrule
    \RC{12}     & \makecell[c]{ReFS Dedup Svc}                                & \dfree        & \medium~\ding{233}~\system & CVE-2025-59210 \\
    \midrule
    \RC{13}     & \multirow{7}{*}{\makecell[c]{Shell\\Infrastructure Host}}   & \uaf          & \low~\ding{233}~\medium    & CVE-2026-20918 \\
    \RC{14}     &                                                             & \uaf          & \low~\ding{233}~\medium    & Confirmed \\
    \RC{15}     &                                                             & \uaf          & \low~\ding{233}~\medium    & Confirmed \\
    \RC{16}     &                                                             & \uaf          & \low~\ding{233}~\medium    & Confirmed \\
    \RC{17}     &                                                             & \uaf          & \low~\ding{233}~\medium    & Confirmed \\
    \RC{18}     &                                                             & \uaf          & \low~\ding{233}~\medium    & Confirmed \\
    \RC{19}     &                                                             & \uaf          & \low~\ding{233}~\medium    & Confirmed \\
    \midrule
    \RC{20}     & \multirow{3}{*}{\makecell[c]{Windows\\Bluetooth Svc}}       & \dfree        & \low~\ding{233}~\system    & CVE-2025-59289 \\
    \RC{21}     &                                                             & \uaf          & \low~\ding{233}~\system    & CVE-2025-53802 \\
    \RC{22}     &                                                             & \uaf          & \low~\ding{233}~\system    & CVE-2025-59220 \\
    \midrule
    \RC{23}     & \multirow{2}{*}{\makecell[c]{Windows\\Digital Media}}       & \uaf          & \low~\ding{233}~\medium    & CVE-2025-53150 \\
    \RC{24}     &                                                             & \uaf          & \low~\ding{233}~\medium    & CVE-2025-59515 \\
    \bottomrule
  \end{tabular}
}

  }
  \label{tab:benchmark-0days}
\end{table}

\subsection{RQ2: Tool Set and Model Contributions}
\label{ss:ablation}

In this RQ, we measure how tool support and model choice affect COTS binary vulnerability discovery and validation.

\PP{Impacts of Tool Support.}
\sys is a \react agent built from scratch with standard components (file-based memory, context management, task verification), and we ask whether different tool support affects its security performance on COTS binaries.
In addition, we supplement production agents with our customized tools to study their performance with enhanced tooling.
Accordingly, we ablate at toolset granularity, not per scaffold component, following the design rationale in \autoref{ss:agent}.
Bare \sys denotes the complete scaffold and toolset, so \sys variants use \wout{\TOOLS} for tool removal.
Bare \codex and \claudecode denote the default binary-navigation setting with only decompilation, disassembly, and symbol lookup, so their variants use \with{\TOOLS} for added \slyptools.
For validation, \SUB denotes the \cc{submit\_poc} execution harness and appears as \with{\SUB} when added to a bare production agent, or \wout{\TOOLS}\with{\SUB} when retained after removing the rest of the toolset from \sys while leaving the default navigation tools.

For vulnerability discovery, as shown in \autoref{tab:rq2-discovery-ablation}, \sys still outperforms production agents even without customized tool support.
Under the same frontier model, \sys detects 1 and 10 more vulnerable functions than \codex and \claudecode, respectively.
Tool support is nonetheless essential for discovery on COTS binaries, and equipping production agents with the full \sys toolset further improves their performance to approach \sys.
For example, with \sys tools, \claudecode detects 62 vulnerable functions, only two fewer than \sys on the same Opus~4.6 model.
We make similar observations for validation (\autoref{tab:rq2-poc-ablation}), where production agents verify 0 cases without the target adapter and debugging, while \syswoslyptools verifies 1--2 cases through its scaffold alone.
Tools are likewise essential for validation, and the \with{\SUB} execution harness plays a key role.
The \with{\SUB} and \wout{\TOOLS}\with{\SUB} rows raise the range to 11--14/40, and the full toolset reaches 22--27/40 across all four evaluated frontier configurations.
Removing tools from \sys also cuts GPT-5.4 discovery runtime by 81.3\% but leaves vulnerable functions uncovered, while adding \slyptools to \codex restores full coverage.
This ablation isolates the coverage contribution of tool support under the same model.

\PP{Impacts of Model Choice.}
To investigate the impact of model choice on security reasoning of COTS binaries, we run \sys with several frontier models for both discovery and validation (\autoref{tab:rq2-ablations}).
\sys remains competitive for discovery when paired with any of the studied models.
For example, \sys still detects 42 vulnerable functions with only 7 false positives even when paired with the weakest Haiku~4.5 model, demonstrating the power of recent frontier models for vulnerability discovery on COTS binaries.
In contrast, different models exhibit larger performance differences on validation.
The weakest Haiku~4.5 verifies only 12.5\% of cases, while Opus~4.6 verifies 67.5\%.
The main reason is that validation requires longer-horizon security analysis (reflected by the token consumption in \autoref{tab:rq2-ablations}), which is beyond the capabilities of weaker models.

\begin{table}[!t]
  \centering
  \caption{Vulnerability discovery ablation on two frontier models. \wout{\TOOLS} removes optional tools from \sys, and \with{\TOOLS} adds those tools to a bare production agent. \textbf{Found} counts recovered vulnerable entry functions out of 64, \textbf{FP} counts false positives, and both are pooled across each object's strongest run. Tool Calls, Tok, and Runtime are per-object averages. Tok is total tokens in millions.}
  \resizebox{\columnwidth}{!}{
    \setlength{\tabcolsep}{3pt}
\begin{tabular}{@{}cc cc ccc@{}}
  \toprule
  & \textbf{Configuration}
  & \makecell{\textbf{Found}\\\textbf{/64}}  & \textbf{FP}
  & \makecell{\textbf{Tool}\\\textbf{Calls}}
  & \makecell{\textbf{Tok}}
  & \makecell{\textbf{Runtime}} \\
  \midrule
  \multirow{4}{*}{GPT-5.4}
  & \cellcolor{gray!15}\sys  & \cellcolor{gray!15}\textbf{64/64} & \cellcolor{gray!15}\textbf{2} & \cellcolor{gray!15}111.7 & \cellcolor{gray!15}3.0M & \cellcolor{gray!15}34.8m \\
  & \syswoslyptools          & 62/64                             & 6                             & 59.3                     & 0.8M                    & 6.5m \\
  & \codex                   & 61/64                             & 5                             & 67.7                     & 1.1M                    & 4.4m \\
  & \codexwithslyptools      & \textbf{64/64}                    & 7                             & 97.8                     & 1.3M                    & 8.1m \\
  \midrule
  \multirow{4}{*}{Opus 4.6}
  & \cellcolor{gray!15}\sys  & \cellcolor{gray!15}\textbf{64/64} & \cellcolor{gray!15}\textbf{4} & \cellcolor{gray!15}58.5  & \cellcolor{gray!15}3.4M & \cellcolor{gray!15}13.8m \\
  & \syswoslyptools          & 59/64                             & 10                            & 57.2                     & 1.7M                    & 10.2m \\
  & \claudecode              & 49/64                             & 17                            & 50.5                     & 3.8M                    & 12.6m \\
  & \claudecodewithslyptools & 62/64                             & 4                             & 58.7                     & 5.0M                    & 12.6m \\
  \bottomrule
\end{tabular}

  }
  \label{tab:rq2-discovery-ablation}
\end{table}

\begin{table}[!t]
  \centering
  \caption{Vulnerability validation ablation on two frontier models over 40 PoC cases. \wout{\TOOLS} removes optional tools from \sys, \with{\TOOLS} adds the full toolset to a bare production agent, and \SUB denotes the \cc{submit\_poc} harness. Filled, half-filled, and open circles indicate full, partial, and no tool availability. Tool Calls and Tok are case averages. Tok is in millions.}
  \resizebox{\columnwidth}{!}{
    \setlength{\tabcolsep}{3pt}
\begin{tabular}{@{}cc ccc cc c@{}}
  \toprule
  & \textbf{Configuration}
  & \textbf{BIN}                    & \textbf{ADP}                   & \textbf{DBG}
  & \textbf{Verified}
  & \makecell{\textbf{Tool}\\\textbf{Calls}}
  & \makecell{\textbf{Tok}} \\
  \midrule
  \multirow{6}{*}{GPT-5.4}
  & \cellcolor{gray!15}\sys         & \cellcolor{gray!15}\fullcircle & \cellcolor{gray!15}\fullcircle & \cellcolor{gray!15}\fullcircle & \cellcolor{gray!15}\textbf{23 / 40} & \cellcolor{gray!15}109.7 & \cellcolor{gray!15}5.7M \\
  & \sys (\wout{\TOOLS}\with{\SUB}) & \halfcircle                    & \opencircle                    & \halfcircle                    & 11 / 40                             & 88.7                     & 1.9M \\
  & \syswoslyptools                 & \halfcircle                    & \opencircle                    & \opencircle                    & 2 / 40                              & 46.0                     & 0.9M \\
  & \codex                          & \halfcircle                    & \opencircle                    & \opencircle                    & 0 / 40                              & 29.0                     & 0.9M \\
  & \codex (\with{\SUB})            & \halfcircle                    & \opencircle                    & \halfcircle                    & 13 / 40                             & 78.7                     & 3.1M \\
  & \codexwithslyptools             & \fullcircle                    & \fullcircle                    & \fullcircle                    & 22 / 40                             & 86.0                     & 5.6M \\
  \midrule
  \multirow{6}{*}{Opus 4.6}
  & \cellcolor{gray!15}\sys         & \cellcolor{gray!15}\fullcircle & \cellcolor{gray!15}\fullcircle & \cellcolor{gray!15}\fullcircle & \cellcolor{gray!15}\textbf{27 / 40} & \cellcolor{gray!15}91.0  & \cellcolor{gray!15}7.7M \\
  & \sys (\wout{\TOOLS}\with{\SUB}) & \halfcircle                    & \opencircle                    & \halfcircle                    & 14 / 40                             & 79.1                     & 4.8M \\
  & \syswoslyptools                 & \halfcircle                    & \opencircle                    & \opencircle                    & 1 / 40                              & 54.9                     & 1.8M \\
  & \claudecode                     & \halfcircle                    & \opencircle                    & \opencircle                    & 0 / 40                              & 55.6                     & 6.8M \\
  & \claudecode (\with{\SUB})       & \halfcircle                    & \opencircle                    & \halfcircle                    & 13 / 40                             & 84.3                     & 12.2M \\
  & \claudecodewithslyptools        & \fullcircle                    & \fullcircle                    & \fullcircle                    & \textbf{26 / 40}                    & 65.9                     & 13.4M \\
  \bottomrule
\end{tabular}

  }
  \label{tab:rq2-poc-ablation}
\end{table}

\begin{table}[!t]
  \centering
  \caption{\sys discovery and validation across six models on the COM benchmark with the full toolset. \textbf{Rate} is verified PoC rate, \textbf{Subs} is total PoC submissions, \textbf{BF} is build-failure rate, \textbf{NC} is no-crash rate after compilation, and \textbf{Tok} is average tokens per case.}
  \resizebox{\columnwidth}{!}{
    \setlength{\tabcolsep}{4pt}
\begin{tabular}{@{}l cc ccccc@{}}
  \toprule
  \multirow{2}{*}{\textbf{Model}}
                & \multicolumn{2}{c}{\textbf{Discovery}}
                & \multicolumn{5}{c}{\textbf{Validation}} \\
  \cmidrule(lr){2-3}\cmidrule(l){4-8}
                & \makecell{\textbf{Found}\\\textbf{/64}} & \textbf{FP}
                & \textbf{Rate (\%)}                      & \textbf{Subs} & \textbf{BF (\%)} & \textbf{NC (\%)} & \textbf{Tok} \\
  \midrule
  GPT-5.4       & \textbf{64}                             & 2             & 57.5             & 129              & 13.2 & 68.2 & 5.6M \\
  GPT-5.3-codex & \textbf{64}                             & 3             & 57.5             & 139              & 8.6  & 74.8 & 6.4M \\
  GPT-5.4-mini  & 59                                      & 3             & 37.5             & 102              & 17.6 & 67.6 & 3.7M \\
  Opus 4.6      & \textbf{64}                             & 4             & \textbf{67.5}    & 61               & 9.8  & 44.3 & 7.7M \\
  Sonnet 4.6    & \textbf{64}                             & 5             & 40.0             & 52               & 30.8 & 38.5 & 9.6M \\
  Haiku 4.5     & 42                                      & 7             & 12.5             & 204              & 27.5 & 70.1 & 2.3M \\
  \bottomrule
\end{tabular}

  }
  \label{tab:rq2-ablations}
\end{table}

\subsection{RQ3: Static COM Race Analysis}
\label{ss:comrace-comparison}

We compare \sys against \comrace~\cite{gu2022comrace}, a static COM race-condition detector.
\comrace identifies cross-references to shared objects between methods without synchronization by tracking field access operations via taint propagation and flagging concurrent accesses without locking primitives.

\PP{Reproduction and Enhancement.}
We develop \comraceplus with three core fixes and four engineering enhancements.
The core fixes are all-register vtable store detection (M1), \cc{this}-pointer propagation (M2), and self-race detection (M3); the enhancements are CFG-based traversal (E1), per-field lock-set tracking (E2), R/R pair filtering (E3), and dereferenced pointer recursion (E4).
We validate \comraceplus against the original \comrace datasets, confirming detection parity with reported capabilities.
M1--M3 are correctness-preserving fixes that allow the tool to produce usable output on production binaries and are therefore always enabled in the Base configuration.
We ablate the heuristic enhancements E1 (CFG traversal), E2 (per-field lock sets), E3 (R/R pair filtering), and E4 (pointer recursion), each of which trades coverage against false positives.

\begin{table}[!tb]
  \centering
  \caption{Per-object outcome of \sys versus \comraceplus on the 20 COM benchmark objects. \textbf{Full} counts objects with every vulnerable function recovered, \textbf{Miss} counts objects with none detected, and \textbf{Partial} covers the rest. \textbf{Found} totals recovered functions out of 64, and \textbf{FP} is the pooled false-positive count. \comraceplus rows add marked enhancements over the always-enabled core (M1--M3).}
  \resizebox{\columnwidth}{!}{
    \setlength{\tabcolsep}{6pt}
\begin{tabular}{@{}l ccc cr@{}}
  \toprule
                                    & \multicolumn{3}{c}{\textbf{Objects}} & \multicolumn{2}{c}{\textbf{Total}} \\
  \cmidrule(lr){2-4} \cmidrule(l){5-6}
  \textbf{System}                   & \textbf{Full}                        & \textbf{Partial}     & \textbf{Miss}                 & \makecell{\textbf{Found}\\\textbf{/64}} & \textbf{FP} \\
  \midrule
  \cellcolor{gray!15}\sys (GPT-5.4) & \cellcolor{gray!15}\textbf{20}       & \cellcolor{gray!15}0 & \cellcolor{gray!15}\textbf{0} & \cellcolor{gray!15}\textbf{64}          & \cellcolor{gray!15}\textbf{2} \\
  \midrule
  \comraceplus (Base)               & 9                                    & 6                    & 5                             & 32                                      & 189 \\
  \comraceplus (+E1/E2)             & 11                                   & 4                    & 5                             & 35                                      & 216 \\
  \comraceplus (+E3/E4)             & 10                                   & 4                    & 6                             & 32                                      & 171 \\
  \comraceplus (+E3)                & 9                                    & 5                    & 6                             & 31                                      & 171 \\
  \comraceplus (+E2/E3/E4)          & 8                                    & 6                    & 6                             & 30                                      & 147 \\
  \bottomrule
\end{tabular}

  }
  \label{tab:rq3-comrace}
\end{table}

\autoref{tab:rq3-comrace} reports per-object outcomes for \sys and every distinct \comraceplus configuration.
\sys with GPT-5.4 fully solves all 20 objects at 2 false positives, whereas no \comraceplus configuration fully solves more than 11 and each leaves at least 5 objects with nothing recovered.
No enhancement combination escapes the underlying tradeoff: the highest-coverage setting (+E1/E2) recovers 35 of 64 functions but emits 216 false positives, while the lowest-noise setting (+E2/E3/E4) cuts false positives to 147 only by recovering 30.
Across the sweep \comraceplus never exceeds 35 recovered functions, so even \sys's weakest model (Haiku~4.5, 42 of 64) surfaces more true vulnerabilities than any static configuration.

\PP{False-Positive and Coverage Analysis.}
\comraceplus's false positives stem from exhaustive pair enumeration: for each \cc{this+offset} field accessed by $n$ methods with inconsistent lock sets, all $n$ are flagged as racing, yet Angr's VEX IR~\cite{shoshitaishvili2016sok} cannot distinguish security-relevant manipulations from benign copies.
The precision enhancements only trade one error for another: adding R/R pair filtering (E3) to Base removes 18 false positives (189$\to$171) but also drops a recovered function (32$\to$31), and the larger reductions toward 147 false positives come at further coverage loss.

The missed functions stem from three structural limitations of static binary analysis on WinRT COM binaries: WinRT's \cc{produce<T,I>} template breaks taint propagation through branchless \cc{this}-pointer adjustments, virtual dispatch through stack-allocated structs requires points-to analysis beyond flat taint tracking, and partially-recovered statically-linked helpers cause missed field accesses.
LLM reasoning recognizes these patterns semantically, recovering all 64 functions.

\subsection{RQ4: Real-World Discovery Beyond the Benchmark}
\label{ss:kernel-extensibility}

RQ4 evaluates whether \sys continues to produce findings beyond the controlled COM benchmark.
We apply the same Stage~1 binary exploration server and Stage~2 compile-execute-debug loop to additional COM services and kernel drivers, changing only prompts for target entry points, harness style, and vulnerability class.
We manually validate all findings, report them to MSRC, and use the results to assess same-surface extension and cross-surface generalization.

\begin{table}[!t]
  \centering
  \caption{Zero-days discovered beyond the benchmark: 7 COM/RPC service bugs (\protect\RC{25}--\protect\RC{31}) and 8 kernel bugs (\protect\RC{32}--\protect\RC{39}). All are confirmed by MSRC. Kernel cases use the same Stage~2 loop with target-specific prompts.}
  \resizebox{\columnwidth}{!}{%
    \setlength{\tabcolsep}{3pt}
{%
  \begin{tabular}{c c c c c}
    \toprule
    \textbf{\#} & \textbf{Target}                                    & \textbf{Type} & \textbf{Sec.~Boundary}      & \textbf{Status} \\
    \midrule
    \multicolumn{5}{c}{\cellcolor{gray!15}\textit{Additional COM/RPC services}} \\
    \RC{25}     & \multirow{2}{*}{\makecell[c]{Clipboard\\User Svc}} & \uaf          & \low~\ding{233}~\medium     & CVE-2026-50384 \\
    \RC{26}     &                                                    & \uaf          & \low~\ding{233}~\medium     & CVE-2026-49183 \\
    \midrule
    \RC{27}     & \makecell[c]{Network Connections}                  & \uaf          & \medium~\ding{233}~\system  & CVE-2026-50476 \\
    \midrule
    \RC{28}     & \makecell[c]{Windows Installer}                    & \uaf          & \medium~\ding{233}~\system  & CVE-2026-50490 \\
    \midrule
    \RC{29}     & \makecell[c]{Work Folders Svc}                     & \uaf          & \medium~\ding{233}~\system  & Confirmed \\
    \midrule
    \RC{30}     & \makecell[c]{Windows\\Management Svc}              & \uaf          & \medium~\ding{233}~\system  & Confirmed \\
    \midrule
    \RC{31}     & \makecell[c]{Network Connection\\Broker Service}   & \dfree        & \medium~\ding{233}~\system  & Confirmed \\
    \midrule
    \multicolumn{5}{c}{\cellcolor{gray!15}\textit{Cross-surface kernel drivers}} \\
    \RC{32}     & \multirow{2}{*}{\makecell[c]{\cc{win32k.sys}}}     & \uaf          & \low~\ding{233}~\system     & CVE-2026-49798 \\
    \RC{33}     &                                                    & \uaf          & \low~\ding{233}~\system     & CVE-2026-33840 \\
    \midrule
    \RC{34}     & \multirow{3}{*}{\makecell[c]{\cc{ntoskrnl.exe}}}   & \oob          & \sandbox~\ding{233}~\system & CVE-2026-32195 \\
    \RC{35}     &                                                    & \oob          & \medium~\ding{233}~\system  & Confirmed \\
    \RC{36}     &                                                    & \uaf          & \sandbox~\ding{233}~\system & Confirmed \\
    \midrule
    \RC{37}     & \makecell[c]{\cc{afd.sys}}                         & \oob          & \sandbox~\ding{233}~\system & Duplicated \\
    \midrule
    \RC{38}     & \multirow{2}{*}{\makecell[c]{\cc{Http.sys}}}       & \ovf          & \medium~\ding{233}~\system  & Confirmed \\
    \RC{39}     &                                                    & \ovf          & \medium~\ding{233}~\system  & Confirmed \\
    \bottomrule
  \end{tabular}
}

  }
  \label{tab:rq4-0days}
\end{table}

\PP{Additional COM Discovery}
Beyond the 20 benchmark objects, \sys discovers 7 additional COM/RPC service zero-days across 6 further production services, as shown in \autoref{tab:rq4-0days}, comprising 6 \uaf bugs and 1 \dfree bug with 2 \low-to-\medium and 5 \medium-to-\system boundaries.
These findings bring \sys's COM/RPC yield in \autoref{ss:default-comparison} to 31 vulnerabilities across twelve services without pipeline changes, confirming continued zero-day discovery on the same surface.

\PP{Cross-Surface Generalization to the Kernel}
The cross-surface test is whether the same cores transfer to a fundamentally different attack surface.
Kernel drivers have no CLSID activation or COM interface metadata, so the target adapter is simply not invoked, while the binary exploration and dynamic debugging cores carry over verbatim.
Driven only by a kernel-specific prompt, \sys produces 8 PoC-verified kernel zero-days spanning 4 driver components (\cc{win32k.sys}, \cc{ntoskrnl.exe}, \cc{afd.sys}, \cc{Http.sys}) and 3 vulnerability classes (3 \uaf, 3 \oob, 2 \ovf).
The security boundaries also broaden beyond the COM benchmark to include \sandbox-to-\system.
This result confirms that \sys is generalizable to new target surfaces and bug classes without new tooling.

\subsection{Threats to Validity}
\label{s:threats}

\PP{Internal Validity}
One threat to internal validity lies in potential data leakage. All zero-day patches in our dataset post-date the latest model knowledge cutoff (Aug. 2025), so the zero-day cases are free from training-data leakage.
The 8 1-day objects may predate the cutoff, but leakage remains impractical: Microsoft ships fixes as cumulative binary updates without the field offsets, object layouts, and synchronization patterns the task requires, and \sys operates on decompiled binaries, not source code or advisories.

\PP{External Validity}
Our controlled evaluation centers on one platform, COTS Windows binaries, so the findings may not directly generalize to other COTS ecosystems.
Meanwhile, Windows is one of the most widely deployed and complicated platforms, and has also been widely studied by prior work on traditional COTS binary analysis~\cite{gu2022comrace,zhang2023comfusion,han2023queryx,kim2017cab}.
In addition, besides our main study on Windows COM services, we have also covered Windows kernel-level binaries to demonstrate the generalizability of our findings.

\section{Related Work}
\label{s:relwk}

\PP{Agentic LLM Scaffolds.}
\react~\cite{yao2022react} establishes tool-interleaved reasoning, a pattern adopted by SWE-agent~\cite{yang2024swe}, OpenHands~\cite{wang2024openhands}, \codex~\cite{codex2025}, and \claudecode~\cite{claudecode2026}.
Long-context degradation remains a shared challenge~\cite{liu2024lost}, motivating summarization and compression strategies~\cite{openhands2025condense,xiao2025improving}.
\sys uses the same \react paradigm but adds checkpointed compaction and task verification for long-horizon binary analysis.

\PP{Agentic Vulnerability Discovery.}
Recent security agents cover source-code vulnerability discovery~\cite{projectzero2024naptime,projectzero2024bigsleep}, CTF-style tool use~\cite{abramovich2025enigma}, checker synthesis and AIxCC-style pipelines~\cite{yang2025knighter,kim2025atlantis}, concolic execution~\cite{luo2026agentic}, and general agentic binary analysis~\cite{chen2025clearagent}.
Automated PoC systems target binary CTF challenges, open-source bugs, C/C++ vulnerabilities, and 1-day web vulnerabilities~\cite{peng2025pwngpt,nitin2025faultline,zhao2026anypoc,fang2024llm}.
\sys instead evaluates live COTS binary services, using reusable MCP tools, target adapters, and debugger-driven PoC repair. For COM races, it uses the target domain studied by \comrace but replaces static rule matching with LLM orchestration over MCP tools and live debugging. For kernel targets, it drops the adapter and reuses the same binary exploration and debugging cores.

\PP{Binary Analysis and LLM-Assisted Reverse Engineering.}
Windows COM and C++ binary analyses include \comrace's taint and lock-set race detector~\cite{gu2022comrace}, \comfusion's union type-confusion analysis~\cite{zhang2023comfusion}, and class or virtual-dispatch recovery systems such as OOAnalyzer~\cite{schwartz2018ooanalyzer}, MARX~\cite{pawlowski2017marx}, and CALLEE~\cite{zhu2023callee}.
LLM-assisted binary work improves decompiler output~\cite{hu2024degpt}, recovers variables and structs~\cite{xie2024resym}, predicts function names~\cite{xu2025gennm}, supports recompilable decompilation~\cite{tan2024llm4decompile,wong2025decllm}, and generates taint rules for IoT firmware~\cite{liu2025llm}.
Source-level graph systems such as LLMxCPG~\cite{lekssays2025llmxcpg} rely on ASTs and type information that production binaries lack, whereas \sys operates on decompiler output and resolves virtual dispatch through vtable analysis.

\PP{Race Condition and Concurrency Bug Detection.}
Traditional fuzzing~\cite{oss-fuzz}, LLM-driven fuzzing~\cite{deng2023large,xia2024fuzz4all}, deterministic kernel race fuzzing~\cite{jeong2019razzer}, and binary-level sanitization~\cite{schilling2024binarytsan} face narrow race windows, privileged harnesses, missing source, or lost type annotations. A survey confirms that data races remain difficult to cover~\cite{upadhyay2023navigating}.
\comrace~\cite{gu2022comrace} is our direct baseline, detecting races by static taint propagation and lock-set analysis on VEX IR.
\sys uses the same problem setting for controlled comparison, but contributes an agentic workflow that toolizes binary inspection, reasons semantically, repairs PoCs under a debugger, and reuses the same cores on kernel targets.

\section{Conclusion}
\label{s:concl}

We present \sys as an end-to-end system for agentic vulnerability reasoning on COTS Windows binaries, combining binary exploration, optional target adapters, and debugger-guided PoC validation.
On the 20-object COM benchmark, \sys recovers all 64 vulnerable entry functions with 2 false positives, verifies PoCs for 67.5\% of cases while production agents verify none, and exceeds the best \comraceplus configuration, which recovers at most 35 functions.
Full-tool ablations show that reusable tools materially affect COTS binary reasoning, while prompt-only adaptation extends \sys to kernel targets and produces 8 PoC-verified kernel zero-days.
Across the campaign, \sys reports 31 COM/RPC service zero-days and 8 kernel zero-days to MSRC, yielding 23 CVEs and \$203{,}000 in bounties.

\bibliographystyle{IEEEtran}
{\footnotesize\raggedright
\bibliography{p,conf}}

\end{document}